\documentclass[aps,twocolumn,showpacs,preprintnumbers,amsmath,amssymb]{revtex4}

\usepackage{graphicx}
\usepackage{dcolumn}
\usepackage{bm}
\usepackage{txfonts}
\usepackage{lscape}
\begin{document}

\title{Incommensurate antiferromagnetic fluctuations in the two-dimensional Hubbard model}

\author{H. C. Krahl}  
\author{S. Friederich}  
\author{C. Wetterich}

\affiliation{\mbox{\it Institut f{\"u}r Theoretische Physik,
Universit\"at Heidelberg,
Philosophenweg 16, D-69120 Heidelberg, Germany}}

\begin{abstract}
Commensurate and incommensurate antiferromagnetic fluctuations in the two-dimensional repulsive $t-t'$-Hubbard model are investigated using functional renormalization group equations. For a sufficient deviation from half filling we establish the existence of local incommensurate order below a pseudocritical temperature $T_{pc}$. Fluctuations not accounted for in the mean field approximation are important---they lower $T_{pc}$ by a factor $\approx2.5$.
\end{abstract}



\maketitle
The two-dimensional Hubbard model \cite{hubbard,kanamori,gutzwiller} has attracted much interest in the past two decades because it is a candidate model for the $\rm{CuO}_2$-planes in the high $T_c$-cuprates and may exhibit d-wave superconducting order \cite{anderson,scalapino} at finite chemical potential. The model shows other interesting order structures such as incommensurate antiferromagnetism which appears close to half filling.

We focus on repulsive interactions $U$ and not too large next-to-nearest neighbour hopping $t'$, where the model is an antiferromagnet at half filling. Not so far away from half-filling a more complicated form of antiferromagnetism, namely incommensurate antiferromagnetism is suggested by mean field computations and numerical studies for finite systems \cite{zaanen,machida,schulz,kato,benard,chubukov,mancini,kaga,moreo,bulut}. Incommensurate antiferromagnetism is related to the existence of spiral magnetic states which occur at large values of $U$ \cite{sarker,zhou,wiese}. Experimentally, incommensurate antiferromagnetism manifests itself in the peak structure of the magnetic structure factor which is accessible via neutron-scattering. It has been observed for a variety of high $T_c$-cuprates, for experimental and numerical results see \cite{birgeneau,cheong,sternlieb,yuan,lorenzanaseibold,fujita,seiboldlorenzana}.

In the temperature region where local incommensurate antiferromagnetic order supposedly sets in, the effective interaction between the electrons is large such that perturbative methods are not reliable. Collective fluctuations of electron-hole pairs in the antiferromagnetic channel play an important role. Since they are omitted in a mean field treatment one may doubt whether the mean field results for incommensurate antiferromagnetism are reliable. For these reasons we investigate the issue of incommensurate antiferromagnetism by a method that is intrinsically non-perturbative and includes effective collective bosonic fluctuations, namely the functional renormalization group for the ``flowing action'' (or ``average action'') $\Gamma_k$ \cite{cw93,berges_review02}. For this scale dependent effective action (or coarse grained free energy) the scale $k$ indicates an infrared cutoff such that only fluctuations with momenta larger than $k$ are effectively included. (Finally, one is interested in the limit $k\rightarrow0$, where $\Gamma_{k\rightarrow0}$ equals the effective action---the generating functional of 1PI-correlation functions---including all fluctuations.) We will work in a version where the dominant collective bosonic fluctuations are represented by bosonic fields \cite{bbw04,bbw05}. Our model is equivalent to the purely fermionic Hubbard model from which it is derived by means of a Hubbard-Stratonovich transformation \cite{hubbardtransf,stratonovich}. Earlier studies employing the present framework have focused on the temperature dependence of \textit{commensurate} antiferromagnetic order \cite{bbw04}, the Kosterlitz-Thouless transition in a more general class of Hubbard-type models \cite{kw07}, and the generation of a coupling in the $d$-wave superconducting channel \cite{krahlmuellerwetterich}. The role of incommensurate antiferromagnetic fluctuations was not taken into account in this earlier work. Functional renormalization group treatments of the Hubbard model are more often given in a purely fermionic formulation, see \cite{zanchi1,halbothmetzner,halbothmetzner2,honerkamp01,salmhofer,honerkampsalmhofer01}. Ref. \onlinecite{halbothmetzner2} is of particular interest since, in accordance with the results described here, it also reports on a region in the phase diagram where incommensurate spin density fluctuations dominate. The present paper also includes an independent computation of the size of the incommensurability that occurs.

Our ansatz for the flowing action includes contributions for the electrons, for the bosons in both the antiferromagnetic and d-wave superconducting channels, and for interactions between fermions and bosons:

\begin{eqnarray}\label{eq:simplesttruncation}
\Gamma_k[\chi]=\Gamma_{F,k}[\chi]+\Gamma_{\mathbf a,k}[\chi]+\Gamma_{F\mathbf a,k}[\chi]+\Gamma_{d,k}[\chi]+\Gamma_{Fd,k}[\chi]
\,.\end{eqnarray}
The collective field $\chi=(\mathbf{a},d,d^*,\psi,\psi^*)$ describes fermion fields $\psi,\psi^*$, the ``antiferromagnetic boson field'' $\mathbf{a}$ and the complex field $d$ a finite expectation value of which signals $d$-wave superconductivity.
The fermionic kinetic term 
\begin{eqnarray}\label{eq:fermprop}
\Gamma_{F,k}=\sum_{Q}\psi^{\dagger}(Q)P_F(Q)\psi(Q)
\end{eqnarray}
involves the inverse fermion propagator
\begin{eqnarray}\label{eq:PF}
P_{F}(Q)=Z_F(i\omega+\xi(\mathbf q))
\,,\end{eqnarray}
where $\xi(\mathbf q)=-\mu-2t(\cos q_x +\cos q_y)-4t' \cos q_x\cos q_y$ depends on the chemical potential $\mu$ and the nearest and next-to-nearest neighbor hopping parameters $t$ and $t'$ of the Hubbard model. We employ a compact notation $X=(\tau,\mathbf x)$, $Q=(\omega,\mathbf{q})$,
\begin{eqnarray}\label{eq:sumdefinition}
\sum\limits_X=\int\limits_0^\beta d\tau\sum\limits_{\mathbf{x}},\quad\sum\limits_Q=T\sum\limits_{n=-\infty}^\infty \int\limits_{-\pi}^\pi \frac{d^2q}{(2\pi)^2}\,,\nonumber\\
\delta(X-X')=\delta(\tau-\tau')\delta_{\mathbf{x},\mathbf{x'}}\,,\hspace{1.5cm}\nonumber\\
\delta(Q-Q')=\beta\delta_{n,n'}(2\pi)^2\delta^{(2)}(\mathbf{q}-\mathbf{q'})\,.\hspace{1cm}
\end{eqnarray}
where all components of $X$ or $Q$ are measured in units of the lattice distance $\mathrm a$ or $\mathrm{a}^{-1}$.
The discreteness of the lattice is reflected by the $2\pi$-periodicity of the momenta $\mathbf{q}$. A scale dependent fermionic wave function renormalization $Z_F$ is included in Eq. \eqref{eq:fermprop}.

The purely antiferromagnetic bosonic term is described by a kinetic term and a local effective potential
\begin{eqnarray}
\Gamma_{a,k}
       =\frac{1}{2}\sum_{Q}\mathbf{a}^{T}(-Q)P_{a}(Q)\mathbf{a}(Q) 
                      +\sum_XU_{a,k}[\mathbf{a}]
\,,\end{eqnarray}
where we employ a quartic effective potential $U_{\mathbf a}$ for $\mathbf a$:
\begin{eqnarray}\label{eq:Ueff}
U_{\mathbf a}[\mathbf{a}]&=&\bar{m}^2_{a}\alpha
+\frac{1}{2}\bar\lambda_a\alpha^2,
\,\end{eqnarray}
with $\alpha=\mathbf a^2/2$.
The kinetic term $P_a$ involves the $Q$-dependent part of the inverse antiferromagnetic propagator and therefore contains the essential information about different kinds of magnetism. Our treatment of this term is discussed in detail below. Local antiferromagnetic order in domains of size $k^{-1}$ is signalled by a minimum of $\Gamma_{a,k}$ for $\mathbf{a}(Q)\neq0$. For $Q=0$ this describes commensurate antiferromagnetism, while a minimum for non-vanishing $\mathbf q$ in $Q=(0,\mathbf q)$ indicates incommensurate antiferromagnetism. A Yukawa-like interaction term couples the bosonic field to the fermions,
\begin{eqnarray}\label{eq:GFak}
&\Gamma_{F\mathbf a,k}=-\bar h_a\!\sum_{K,Q,Q'}\delta(K+\Pi-Q+Q')
\\
& \hspace{2.5cm}\times\;
        \mathbf{a}(K)\cdot[\psi^{\dagger}(Q)\boldsymbol{\sigma}\psi(Q')]\nonumber
\,,\end{eqnarray}
where the momentum vector $\Pi$ is given by $\Pi=(0,\pi,\pi)$.


The bosonic field $d$ is associated to Cooper-pairs in the d-wave channel. It is described in more detail in \cite{kw07,krahlmuellerwetterich} where the exact form of $\Gamma_{Fd}$ can be found. In this note we include the effect of d-wave fluctuations on the flow of the fermionic and ``antiferromagnetic'' part of $\Gamma_k$. We use $\Gamma_{d,k}=\sum_{Q} d^{*}(Q)P_d(Q)d(Q)+\sum_XU_{d,k}[d,d^*]$ with $U_{d}[d,d^*]=\bar{m}^2_{d}\delta+\bar\lambda_d\delta^2/2$ where $\delta=d^*d$. Here we focus exclusively on the emergence of (either commensurate or incommensurate) magnetic order which occurs in the vicinity of half-filling. The emergence of $d$-wave superconducting order at larger values of $|\mu|$ will be discussed in detail in a future publication. No superconductivity has been detected in the region of the phase diagram studied here.

The dependence of the flowing action on the scale $k$ is described by an exact flow equation \cite{cw93}. Our ansatz \eqref{eq:simplesttruncation} approximates the solutions to this functional differential equation. At the microscopic scale $k=\Lambda$ the flowing action must be equivalent to the microscopic action of the Hubbard model. Since we want to eliminate the (constant) four-fermion coupling $U$ at $k=\Lambda$ which, of course, has no contributions exhibiting $d$-wave symmetry, the repulsive interaction between the fermions must be contained in the antiferromagnetic Yukawa coupling $\bar h_a$. In the bosonized picture one has, instead of the original four-fermion coupling $U$ a boson-mediated interaction term $\bar h_a^2/\bar{m}_a^2$ which must be chosen proportional to $U$. Since an additional sum over spin directions has to be carried out it has to be chosen as $U/3$. Thereby we have simply transcribed the original model into an equivalent one using bosonic language. Since the original model does not contain any eight electron terms, no quartic bosonic coupling $\bar\lambda_a$ can exist at the UV scale $k=\Lambda$. In sum, a set of possible ``initial conditions'' for the flow of the coupling constants is given by
\begin{eqnarray}\label{eq:initialcond}
&\bar{m}_a^2|_{\Lambda}=U/3\,,\quad\bar{h}_a|_{\Lambda}=U/3\,,\quad\bar\lambda_a|_{\Lambda}=0\,,\quad P_a(Q)|_{\Lambda}=0,\nonumber\\
&\Gamma_{Fd}|_{\Lambda}=0\,,\quad \Gamma_{d}|_{\Lambda}=d^*d,\quad Z_{F}|_{\Lambda}=1
\,.\end{eqnarray}
These values specify the action $\Gamma_\Lambda$ (or, equivalently, the Hamiltonian) at the microscopic level.

At the microscopic scale $\Lambda$, the action for $\mathbf a$ is Gaussian such that $\mathbf a$ can be ``integrated out'' by solving its field equation as a functional of $\psi$. The $d$-boson decouples and becomes irrelevant. This demonstrates that $\Gamma_\Lambda$ indeed coincides with the purely fermionic Hubbard model with repulsive coupling $U>0$.

\begin{figure}[t]
\begin{center}
\includegraphics[width=85mm,angle=0.]{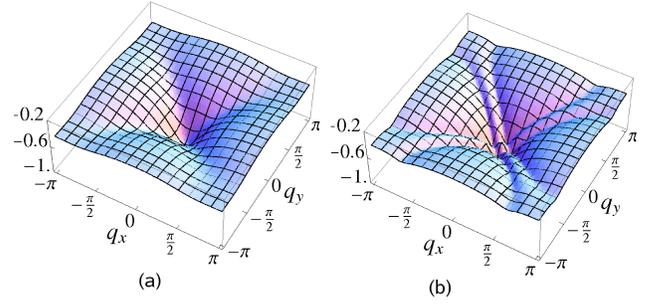}
\end{center}
\caption{\small{Mean field kinetic term $P_{a}(0,\mathbf q)/t$ of the $\mathbf{a}$-boson as a function of space-like momenta for $U/t=3$ and $t'=0$. In Figure (a) $\mu=0$ and $T/t=0.205$, in Figure (b) $\mu/t=-0.27$ and $T/t=0.0435$. Both temperatures are mean field critical temperatures.
}}
\label{fig:oneloopa}
\end{figure}

We still have to specify the truncation for the kinetic term of the $\mathbf a$-boson $P_a(Q)$. This is a central object of this paper, since incommensurate antiferromagnetic fluctuations will dominate if the minimum of $P_a(0,\mathbf q)$ occurs for nonzero $\mathbf q$. In order to gain some first information about the general shape of $P_a$ we compute the mean field contribution from the fermionic loop
\begin{eqnarray}\label{eq:rhoferm}
\Delta P_{a}(Q)=\sum_{P}
\frac{\bar{h}_{a}^2}
{P_F(Q+P+\Pi)P_F(P)}   +(Q\rightarrow -Q)
\,.\end{eqnarray}
The general features of this mean field contribution are used in order to motivate the form of the bosonic propagator in our truncation. We observe close to half filling two qualitatively different situations. At half filling and for sufficiently high temperatures also close to half filling there is a pronounced minimum at $\mathbf{q}=0$, see Fig. \ref{fig:oneloopa} (a). However, away from half filling the picture is different for sufficiently low temperatures, see Fig. \ref{fig:oneloopa} (b). In the center at $\mathbf{q}=0$ there is a local maximum and there are four minima at positions
\begin{eqnarray}\label{eq:positionshatq}
\mathbf{q}_{1,2}=(\pm\hat q,0)\,,\quad \mathbf{q}_{3,4}=(0,\pm\hat q)
\,,\end{eqnarray}
where $\hat q$ is a function of $T$, $\mu$, and $t'$. This is a manifestation of the dominance of {\it incommensurate} antiferromagnetic fluctuations.

Once the minimal value of the inverse bosonic propagator $\left[P_a(0,\mathbf{q})+\bar m_a^2\right]$ at zero frequency becomes smaller than zero, the minimum of the free energy can no longer occur for $\langle\mathbf a(Q)\rangle=0$. One rather expects spontaneous symmetry breaking with a non-zero expectation value of $\langle\mathbf a\rangle$. As long as the minimum of $P_a(0,\mathbf{q})$ is located at $\mathbf{q}=0$, the order parameter $\langle|\mathbf a|\rangle\sim\delta(\mathbf q)$ indicates commensurate antiferromagnetism. However, for a minimum at $\mathbf{q}=\mathbf{q}_j\neq0$ the incommensurate antiferromagnetic order breaks further lattice symmetries. One of the pairs of minima \eqref{eq:positionshatq} is selected and the symmetry of rotations by $\pi/2$ around $\mathbf{q}=0$ in momentum space is spontaneously broken. The spins change sign between neighboring lattice sites only in one direction, the $x$-direction say, whereas in the orthogonal direction the periodicity corresponds to some momentum $\pi\pm \hat q$. The state with $\langle\mathbf a\rangle=0$ becomes unstable when $( P_{a})_{min}=-\bar m_a^2$. In case of a second order phase transition this occurs for the mean field critical temperature $T=T_{MFc}$. Figures \ref{fig:oneloopa} (a) and (b) correspond to mean field critical temperatures. Note that the system selects one of the \textsl{pairs} $\mathbf{q}_{1,2}$ or $\mathbf{q}_{3,4}$ since $\mathbf a(X)$ is a real field. Therefore the system remains symmetric with respect to reflection about the axes.

We are interested in whether incommensurate antiferromagnetism persists if bosonic fluctuations are included. Taking into account bosonic fluctuations, the critical temperature vanishes in the infinite volume limit due to the Mermin-Wagner theorem. The destruction of local order by the long range fluctuations of the Goldstone bosons (antiferromagnetic spin waves) is only a logarithmic effect, however. For a probe of finite macroscopic size antiferromagnetic order can be observed and the critical temperature is nonzero \cite{bbw05}. Here the effective critical temperature $T_c$ is defined such that for $T<T_c$ the typical size of ordered domains exceeds the macroscopic size of the probe $l$. In other words, $\langle \mathbf a(k) \rangle$ differs from zero for $k_{ph}=l^{-1}$ if $T<T_c$, while for $T>T_c$ one has $\langle \mathbf a(k) \rangle=0$.

In this note we only study the pseudocritical temperature $T_{pc}$ which marks the onset of local ordering corresponding to a minimum of the flowing action $\Gamma_k$ for $\mathbf{a}(0,\mathbf q)$. Above this temperature, $\langle|\mathbf a|\rangle=0$ holds on all scales of the renormalization flow. For $T<T_{pc}$ local order sets in for $k=k_c>0$. In case of incommensurate antiferromagnetism we expect the formation of domain walls between regions where $\mathbf{\hat q}$ points in the $x$- or $y$-direction. This constrasts with commensurate antiferromagnetism where only a continuous symmetry is broken for $\mathbf a\neq0$. For $k<k_c$ the flow should then be continued in a regime with nonzero $\mathbf a$ in order to account properly for the Goldstone boson fluctuations. This has been investigated for the commensurate case in \cite{bbw05}, but is not yet implemented for the incommensurate case in the present note. We note that $T_{pc}$ is the equivalent of the mean field critical temperature. For $T_c<T<T_{pc}$ the electron propagator does not exhibit a true gap, but it is suppressed for momenta corresponding to the inverse of length scales for which local order is present.

Inspired by the shape of $P_a$ in the mean field approximation we approximate the kinetic term for the antiferromagnetic boson by
\begin{eqnarray}\label{eq:apropparam}
P_{a,k}(Q)=Z_a\omega^2+A_aF(\mathbf q)\, ,
\end{eqnarray}
where for the case of commensurate antiferromagnetism we choose for $F(\mathbf q)$
\begin{eqnarray}\label{eq:apropparam1}
F_c(\mathbf q)=\frac{D^2[\mathbf{q}]^2}{D^2+[\mathbf{q}]^2}
\,.\end{eqnarray}
Here $[\mathbf{q}]^2$ is defined as $[\mathbf{q}]^2=q_x^2+q_y^2$ for $q_i\in [-\pi,\pi]$ and continued periodically otherwise. For small $\mathbf{q}^2$ the quadratic approximation $P_a=A_a\mathbf{q}^2$ describes a linear dispersion relation for the composite bosonic field, $\omega=\sqrt{A_a/Z_a}|\mathbf{q}|$, while for $\mathbf q$ near the boundary of the Brillouin zone the momentum dependence of $P_a$ `levels off' as in Figs. \ref{fig:oneloopa} (a), (b). For a suitable choice of $A_a$ and $D$ the shape of the mean field result for $P_a$ is well reproduced. Of course, due to the important contributions of bosonic fluctuations beyond mean field theory, the actual values of $A_a$ and $D$ will differ substantially from the mean field values.

Within the functional renormalization group approach, we describe the scale dependence of the bosonic kinetic term by flow equations for the parameters $A_a$ and $D$.  For these purposes we define the gradient coefficient $A_a$ by
\begin{eqnarray}\label{eq:Aa}
A_a=\frac{1}{2}\frac{\partial^2}{\partial l^2}P_a(0,l,0)\big{|}_{l=\hat q}
\end{eqnarray}
with $\hat q=0$ in the commensurate case.
The shape coefficient $D$ is computed as
\begin{eqnarray}\label{eq:D}
D^2=\frac{1}{A_a}\big(P_a(0,\pi,\pi)-P_a(0,\hat q,0)\big).
\end{eqnarray}
The flow equations for $A_a$ and $D$ can be extracted by inserting our truncation in the exact flow equations for the kinetic term \eqref{eq:apropparam}.

\begin{figure}[t]
\begin{center}
\includegraphics[width=60mm,angle=0.]{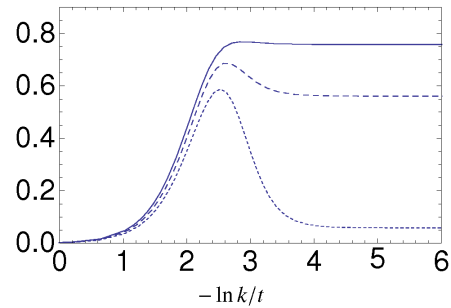}
\end{center}
\caption{\small{Renormalization flow of the gradient coefficient $A_{a,k}$ for $U/t=3$ and $\mu/t=-0.12$ according to Eq. \eqref{eq:flowAa} in the parameter regime where we have commensurate antiferromagnetism and therefore $\partial_k \hat q=0$. The solid line corresponds to $T/t=0.08$, the long-dashed line to $T/t=0.07$, and the short-dashed line to $T/t=0.058$.}}
\label{fig:P_asym}
\end{figure}

During the renormalization flow the gradient coefficient $A_a$ first increases, starting from $A_a=0$ at the scale $\Lambda$. At half filling and in the proximity of half filling for sufficiently high temperatures, $A_a$ either increases monotonically or at least remains larger than zero on all scales $k<\Lambda$, see Fig. \ref{fig:P_asym}. The minimum of $P_a$ occurs for $\mathbf q=0$ and commensurate antiferromagnetic fluctuations dominate.

\begin{figure}[t]
\includegraphics[width=85mm,angle=0.]{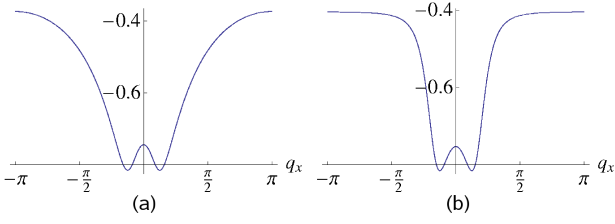}
\caption{\small{
In (a) the mean field approximation for bosonic kinetic term $P_a(0,q_x,0)/t$ is shown as a function of the $x$-component of spatial momenta. Parameters are $U/t=3$, $\mu/t=-0.35$, $t'=0$ and $T/t=0.1$. Fig. (b) shows the same quantity according to our approximation given by Eqs. \eqref{eq:apropparam} and \eqref{eq:apropparaminkomm} with the values of $A_a$, $\hat q$ and $D$ drawn from the mean field computation.}}
\label{fig:menafieldapropmom}
\end{figure}

However, for low enough temperatures and at sufficient distance from half-filling, $A_a$ becomes zero on a certain scale. If we continued to evaluate $A_a$ for $\mathbf q =0$ it would decrease to negative values for lower scales.
This situation corresponds to the case of incommensurate antiferromagnetism. The ansatz for the function $F(\mathbf q)$ given in Eq. \eqref{eq:apropparam1} is no longer suitable. One has to allow for the existence of minima of $P_{a,k}(0,\mathbf{q})$ at nonzero $\mathbf{q}\neq0$.
The ansatz \eqref{eq:apropparam} for the inverse bosonic propagator employs now for $F(\mathbf q)$
\begin{eqnarray}
F_i(\mathbf q,\hat q)&=&\frac{D^2\tilde F(\mathbf q,\hat q)}{D^2+\tilde F(\mathbf q,\hat q)}
\,.\end{eqnarray}
The quadratic momentum dependence of the numerator in \eqref{eq:apropparam1} is replaced by an expression which is quartic in momentum and explicitly includes the incommensurability $\hat q$:
\begin{eqnarray}\label{eq:apropparaminkomm}
\tilde F(\mathbf q,\hat q)=\frac{1}{4\hat q^2}\big((\hat q^2-[\mathbf q]^2)^2+4[q_x]^2[q_y]^2\big)
\,.\end{eqnarray}
The first term in $\tilde F$ vanishes for $[\mathbf q]^2=\hat q^2$ and suppresses the propagator for $[\mathbf q]^2\neq\hat q^2$. The second term favours the minima \eqref{eq:positionshatq} as compared to a situation where rotation-symmetry in the $q_x-q_y$-plane is preserved. The prefactor is determined by Eq. \eqref{eq:Aa}. For $\hat q\rightarrow0$ one has $A_a\sim\hat q^2$ such that $P_a$ becomes quartic in $\mathbf q$. We compare in Fig. \ref{fig:menafieldapropmom} the kinetic term $P_a(0,\mathbf{q})$ in mean field theory with the approximation from our ansatz which shows satisfactory agreement.

\begin{figure}[h]
\begin{center}
\includegraphics[width=60mm,angle=0.]{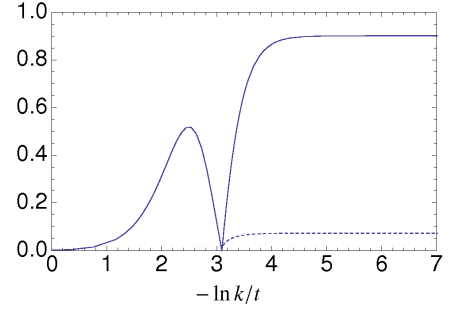}
\end{center}
\caption{\small{Renormalization flow of the gradient coefficient $A_{a,k}$ and the incommensurability $\hat q$ according to Eqs. \eqref{eq:flowAa} and \eqref{eq:calcqhat} at $\mu/t=-0.12$ and $T/t=0.05$, where incommensurate antiferromagnetism dominates. The solid line shows $A_a$ decreasing to zero at $-\ln k/t=3.09$ and increasing again when the incommensurability $\hat q$ (short-dashed line) sets in.}}
\label{fig:P_assb}
\end{figure}

In Fig. \ref{fig:P_assb}, a typical flow for $A_a$ and $\hat q$ in the incommensurate regime is displayed. For scales below the scale where $A_a$ becomes zero, $\hat q$ increases to a finite value and $P_a(0,\mathbf{q})$ has four degenerate minima at positions given by Eq. (\ref{eq:positionshatq}). The solution $\hat q$ of Eq. (\ref{eq:calcqhat}) at the end of the flow corresponds to the position of the minimum, e.g., at the positive $q_x$-axis .
We next specify the flow in more detail.

The regulator function $R^a_k(Q)$ for the antiferromagnetic fluctuations should be adapted in order to allow for the dominance of incommensurate antiferromagnetism. We employ, similarly for the commensurate and incommensurate case,
\begin{eqnarray}\label{regulator}
R^a_k(Q)=A_{a}\cdot(k^2-F_{c,i}(\mathbf q,\hat q))\Theta(k^2-F_{c,i}(\mathbf q,\hat q))
\,,\end{eqnarray}
respectively. This generalizes the cutoff chosen in \cite{krahlmuellerwetterich}.

The flow equation for the gradient coefficient is obtained by taking appropriate derivatives in one of the minima
\begin{eqnarray}\label{eq:flowAa}
\partial_kA_a&=&\sum_Q\bar{h}_a^2\tilde\partial_k\frac{\partial^2}{\partial l^2}\frac{1}{P^k_F(Q)P^k_F(K+Q+\Pi)}\Bigg{|}_{l=\hat q}\nonumber\\
&+&\sum_Q\bar{h}_a^2(\partial_k\hat q)\frac{\partial^3}{\partial l^3}\frac{1}{P^k_F(Q)P^k_F(K+Q+\Pi)}\Bigg{|}_{l=\hat q}
\,,\end{eqnarray}
where $K=(0,l,0)$. We have defined $P_{F,k}(Q)=P_F(Q)+R_k^F(Q)$, with fermion cutoff $R_k^F$ chosen as in \cite{krahlmuellerwetterich}. The first term in \eqref{eq:flowAa} results from the change of the infrared cutoff in the fluctuations. The symbol $\tilde\partial_k$ means a formal derivative with respect to the cutoff function $R_k^F$. The second term in  \eqref{eq:flowAa} reflects the shift of the location of the minimum of $P_a$ at $(\hat q,0)$ and is absent if commensurate fluctuations dominate, $\hat q=0$.

 A flow equation for the position of the minima $\hat q$ is derived from the condition
\begin{eqnarray}
\frac{\partial}{\partial q_x}P_{a,k}(0,\mathbf q)\big{|}_{\mathbf q=(\hat q,0)}=0
\,.\end{eqnarray}
Taking the scale derivative of this equation one obtains the flow equation:
\begin{eqnarray}\label{eq:calcqhat}
(\partial_k\hat q)\frac{\partial^2}{\partial q_x^2}P_{a,k}(0,\mathbf q)\big{|}_{\mathbf q=(\hat q,0)}+\frac{d}{dk}\Big{|}_{\hat q}\frac{\partial}{\partial q_x}P_{a,k}(0,\mathbf q)\big{|}_{\mathbf q=(\hat q,0)}& &\\
=(\partial_k\hat q)2A_a+\frac{d}{dk}\Big{|}_{\hat q}\frac{\partial}{\partial q_x}P_{a,k}(0,\mathbf q)\big{|}_{\mathbf q=(\hat q,0)}&=&0\nonumber
\,.\end{eqnarray}

Flow equations for the other running couplings $Z_F,\bar m_a^2,\bar \lambda_a,\bar h_a,\bar m_d^2,\bar \lambda_d,\bar h_d,D$ are not given explicitly here, see \cite{krahlmuellerwetterich}.

\begin{figure}[h]
\includegraphics[width=80mm,angle=0.]{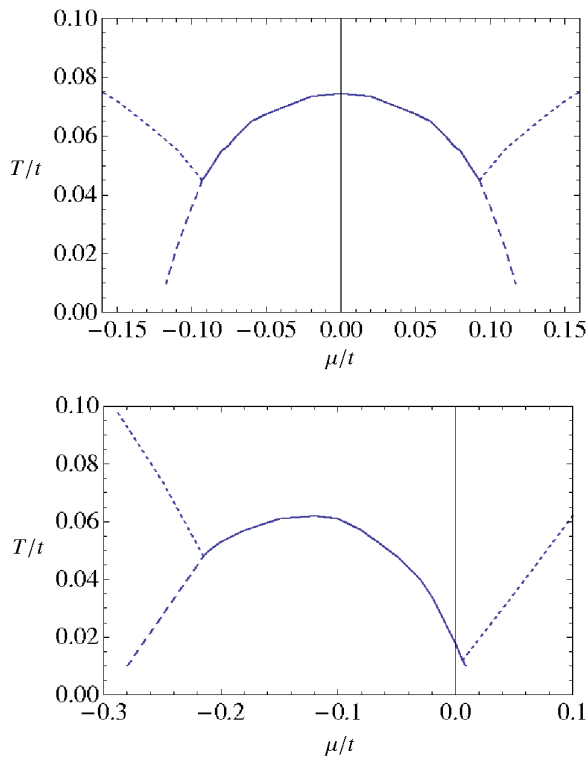}
\caption{\small{Renormalization group results for the pseudocritical temperature $T_{pc}/t$ as a function of $\mu/t$, given by the solid (commensurate) and dashed (incommensurate) lines. Results are displayed for $U/t=3$,  $t'=0$ (upper panel) and $t'/t=-0.05$ (lower panel).}}
\label{fig:phasediagronlyantif}
\end{figure}

We now turn to the results obtained in our renormalization group scheme. An overview of the occurence of incommensurate antiferomagnetism is given in Fig. \ref{fig:phasediagronlyantif}, showing pseudocritical temperatures $T_{pc}$ for the different kinds of antiferromagnetic order in the presence of vanishing (upper panel) and nonvanishing (lower panel) next-to-nearest neighbor hopping $t'$. The solid line signals the onset of local commensurate, the long-dashed line the onset of local incommensurate antiferromagnetic order. Below the short-dashed line there is no local magnetic order but incommensurate fluctuations dominate. Below the point where the short-dashed line terminates at low temperatures, numerical solutions to the flow equations, as we have implemented them numerically, are no longer reliable. For both vanishing and non-vanishing $t'$, one observes commensurate antiferromagnetism for a certain range of chemical potential $\mu$, while for smaller and larger values of $\mu$ incommensurate fluctuations begin to dominate. For finite $t'$, however, the pseudocritical curve is no longer the same for positive and negative $\mu$ but, for negative $t'$, is shifted to more negative values of $\mu$.

The pseudocritical temperature is found to be substantially lower than according to the mean field computation. For $U=3t$, $t'=0$ and $\mu=0$, for example, the mean-field computation gives $T_{MFc}/t=0.205$, while we find $T_{pc}/t=0.0745$ when one takes into account bosonic fluctuations. By reducing the interaction, the shape of the pseudocritical curve remains the same but local order emerges only at lower temperatures. 

\begin{figure}[h]
\includegraphics[width=65mm,angle=0.]{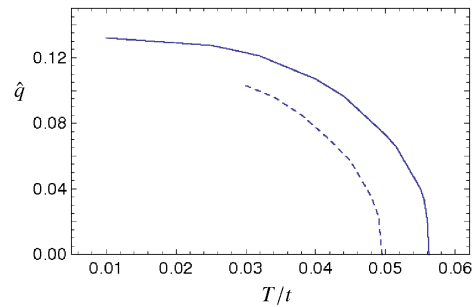}
\caption{\small{Renormalization group results for the incommensurability $\hat q$ as a function of $T$ for $U/t=3$, $t'=0$ and $\mu/t=-0.12$ (solid line) and $\mu/t=-0.105$ (dashed line).}}
\label{fig:qhat}
\end{figure}

With decreasing temperature the tendency towards incommensurate fluctuations is increased, which can be demonstrated by studying the dependence of $\hat q$ on $T$ at fixed chemical potential. It is shown for $\mu/t=-0.105$ and $\mu/t=-0.12$ in Fig. \ref{fig:qhat}. For large enough temperatures one has $\hat q=0$, while below some $\mu$-dependent temperature incommensurate antiferromagnetism sets in. The temperature where this happens is indicated by the short-dashed line in Fig. \ref{fig:phasediagronlyantif} (upper panel). For smaller $T$ the value of $\hat q$ increases, the final point of the $\mu/t=-0.105$-curve at low temperature corresponds to the long-dashed line in Fig. \ref{fig:phasediagronlyantif}

As one can see from the curve representing $\mu/t=-0.12$ in  Fig. \ref{fig:qhat}, at small temperatures the size of the incommensurability is approximately constant. Therefore we compare our result to the zero-temperature result obtained by \cite{schulz} saying that $\hat q=2\arcsin (|\mu|/2t)$ (which has also been used in the fermionic RG computation given in \cite{halbothmetzner2}). For $\mu/t=-0.12$ this formula gives $\hat q\approx0.120$ whereas we find $\hat q\approx0.132$. By taking into account fluctuations the incommensurability seems to be slightly enhanced. Agreement with the results displayed in \cite{mancini} obtained by means of the composite operator method is also satisfactory.

\begin{figure}[h]
\includegraphics[width=85mm,angle=0.]{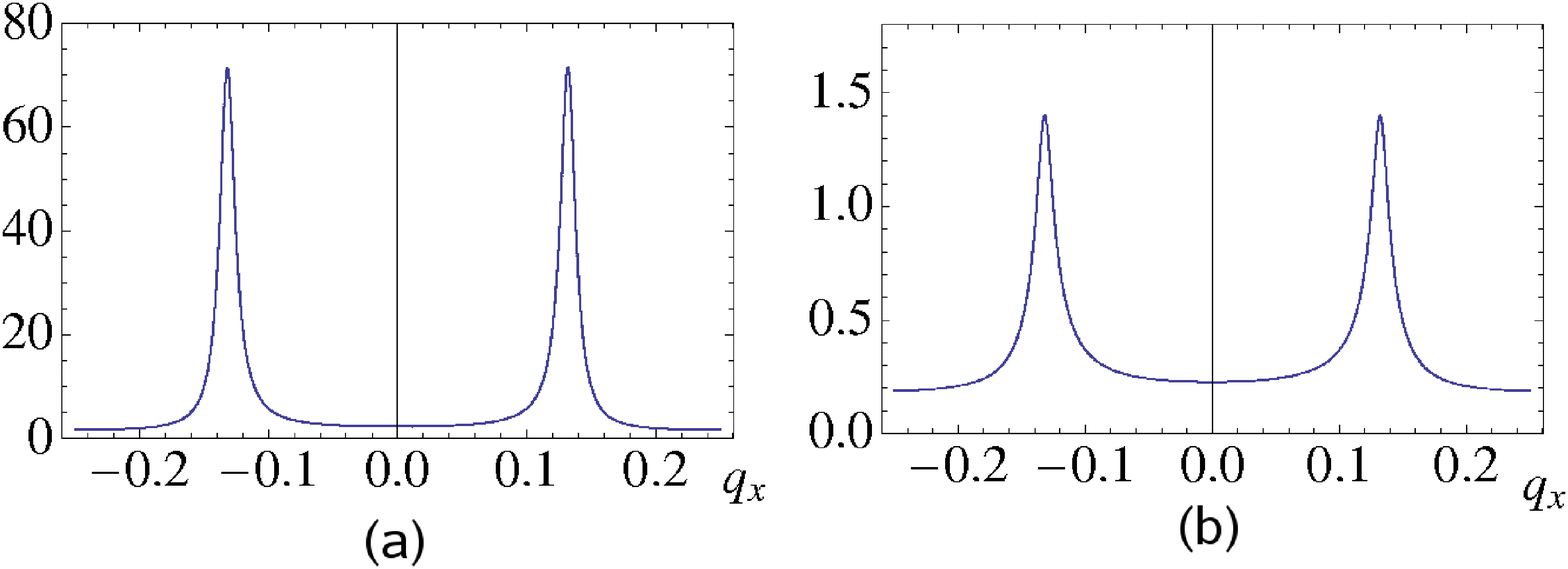}
\caption{\small{Fig. (a) shows the spin susceptibility $(P_a(0,q_x,0)/t)^{-1}$ and Fig. (b) the bosonic occupation number $n_a(q_x,0)$ for $\mu/t=-0.12$, $T/t=0.01$, $U/t=3$ and $t'=0$ according to the renormalization group computations. Both curves are given as a function of spatial momentum in $x$-direction. A peak at $q_x=0$ would signal the dominance of commensurate antiferromagnetism. The actual peaks, located at $\hat q=\pm0.132$, indicate incommensurate antiferromagnetism.}}
\label{fig:suscocc}
\end{figure}

A dominance of incommensurate antiferromagnetic fluctuations can be observed in the momentum dependence of the magnetic susceptibility and the bosonic occupation number. The susceptibility is given by the bosonic propagator at zero frequency $P_a^{-1}(0,\mathbf q)$, while the occupation number is obtained by an additional sum over bosonic Matsubara frequencies, $n_a(\mathbf q)=T\sum_{\omega_B}(P_a(\omega_B,\mathbf q))^{-1}$. Fig. \ref{fig:suscocc} shows that for parameters where the bosonic mass is small, here $\bar m_a^2/U\approx10^{-2}$, and thus close to the onset of local incommensurate order, both the magnetic susceptibility and the bosonic occupation number are peaked at $q_x=\pm\hat q$, signalling that incommensurate fluctuations strongly dominate. The situation is completely analogous for the $q_y$-dependence of the susceptibility at $q_x=0$, whereas both quantities do not have such a pronounced peak structure along the Brillouin zone diagonal.

\begin{figure}[h]
\includegraphics[width=85mm,angle=0.]{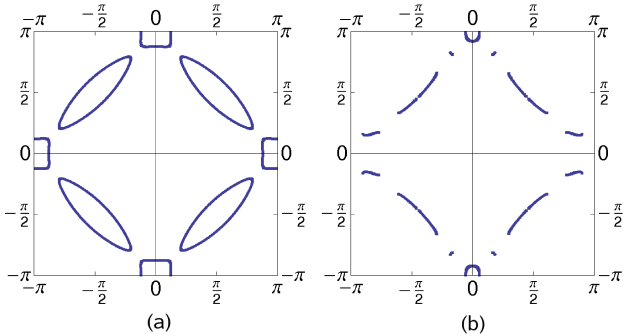}
\caption{\small{Mean field effective Fermi surfaces for $\mu/t=-0.6$, $t'/t=-0.2$ and gap parameter $A/t=0.1$. Fig. (a) shows the commensurate case $\hat q=0$ where the Fermi surface exhibits hole- and particle pockets at the magnetic Brillouin zone boundary. In Fig. (b), the remainders of the effective Fermi surface are shown for a nonzero incommensurability $\hat q=0.3$ along the $x$-axis.}}
\label{fig:incommfermi}
\end{figure}

In those regions of the phase diagram in which (either commensurate or incommensurate) antiferromagnetic order exists on a certain legth scale $k$ our truncation becomes inapplicable in the regime below $k$. The simplest way of obtaining a glimpse at these regimes is by means of a mean field analysis, so before closing the discussion we briefly address this problem. A more extensive mean field treatment, if only with regards to the \textit{commensurate} case but including a nonzero next-to-nearest neighbor hopping $t'$, is given in \cite{reiss}. Here one has to take into account that the periodicity of a system in the N\'eel state is changed resulting in a new ``magnetic'' Brillouin zone whose boundaries are given by the lines between the $(\pm\pi,0)$ and $(0,\pm\pi)$ points. Correspondingly, the mean field dispersion relation for a nonzero gap parameter $A=\bar h_a\langle |\mathbf{a}|\rangle$ has two branches
\begin{eqnarray}\label{eq:dispcomm}
E_\pm(\mathbf p)=\frac{1}{2}\left( \xi(\mathbf{p})+\xi(\mathbf{p}+\boldsymbol \pi)\pm\sqrt{(\xi(\mathbf{p})-\xi(\mathbf{p}+\boldsymbol \pi))^2+4 A^2} \right)
\,\end{eqnarray}
which, for finite $t'$, lead to an interestingly structured effective Fermi surface enclosing hole pockets around $(\pm\pi/2,\pm\pi/2)$ and electron pockets around $(\pm\pi,0)$ and $(0,\pm\pi)$, see the example drawn in Fig. \ref{fig:incommfermi} (a), for further details see e.\,g. \cite{reiss}.

In the presence of a nonzero expectation value $\langle\mathbf{a}(\mathbf{\hat q})\rangle$ with $\mathbf{\hat q}\neq0$, i.\,e. in the presence of incommensurate order, the inverse of the fermionic mean field propagator at zero frequency has contributions from Eqs. \eqref{eq:PF}(with $Z_F=1$) and \eqref{eq:GFak} and is given by
\begin{eqnarray}
&&P_F(\mathbf q, \mathbf{q}')=\xi(\mathbf q)\delta(\mathbf q-\mathbf q')\\&&\hspace{0.3cm}-\frac{\mathbf A\cdot\boldsymbol\sigma}{\sqrt{2}}\left( \delta(\mathbf q-\mathbf q'-\boldsymbol\pi+\mathbf{\hat q})+\delta(\mathbf q-\mathbf q'-\boldsymbol\pi-\mathbf{\hat q}) \right)\nonumber
\end{eqnarray}
with $\mathbf{\hat q}=\mathbf{q}_{1,2}$ or $\mathbf{\hat q}=\mathbf{q}_{3,4}$ as defined in Eq. \eqref{eq:positionshatq}. The analogue of the Fermi surface corresponds to the zero eigenvalues of $P_F$. However, the corresponding eigenmodes are no longer momentum eigenstates. Nevertheless, if the gap parameter $A=|\mathbf A|$ is nonzero but small, many eigenvalues of $P_F(\mathbf q, \mathbf{q}')$ have most of their support each at a single momentum $\mathbf p$. This concerns all those momenta $\mathbf p$ for which the condition 
\begin{equation}\label{eq:cond}
A\ll|\xi(\mathbf{p}+\boldsymbol \pi+\mathbf{\hat q})|,|\xi(\mathbf{p}+\boldsymbol \pi-\mathbf{\hat q})|
\end{equation}
is fulfilled. With respect to these momenta the equation
\begin{eqnarray}\label{eq:disp}
\xi(\mathbf{p})-\frac{A^2}{2}\left(\frac{1}{\xi(\mathbf{p}+\boldsymbol \pi+\mathbf{\hat q})}+\frac{1}{\xi(\mathbf{p}+\boldsymbol \pi-\mathbf{\hat q})}\right)=0
\,\end{eqnarray}
defines an effective Fermi surface which is obtained by (approximately) diagonalizing $P_F(\mathbf q, \mathbf{q}')$ for small $A$. For large enough $A$ the effective Fermi surface vanishes completely because the number of solutions to Eq. \eqref{eq:disp} that satisfy the condition \eqref{eq:cond} rapidly goes down. In Fig. \ref{fig:incommfermi} (b) the effective Fermi surface is shown for the incommensurate case with an order parameter $\langle\mathbf{a}(\mathbf{\hat q})\rangle$ where $\mathbf{\hat q}=\mathbf{q}_{1,2}$, i.\,e. the incommensurability is along the $x$-axis. The symmetry of rotations by $\pi/2$ is manifestly broken.


To summarize, we have shown that incommensurate antiferromagnetic order in the two-dimensional Hubbard model persists if bosonic fluctuations are taken into account. This phenomenon occurs at least in the form of local order for temperatures smaller than the pseudocritical temperature shown in Fig. \ref{fig:phasediagronlyantif}. We speculate that for $T\rightarrow0$ the size of the incommensurate domains grows beyond the size of typical macroscopic probes, but this remains to be shown. If magnetic fluctuations play a role in the generation of d-wave superconducting order, the effect of incommensurability has to be taken into account.

{\bf Acknowledgments}: HCK acknowledges financial support by the DFG research unit FOR 723 under the contract WE 1056/9-1. SF acknowledges support by the Studienstiftung des Deutschen Volkes.

\renewcommand{\thesection}{}
\renewcommand{\thesubsection}{A{subsection}}
\renewcommand{\theequation}{A \arabic{equation}}


\end{document}